\global\def\draftcontrol{0}

   \def\versionno{NP quant geometry III}

\catcode`\@=11

\expandafter\ifx\csname draftcontrol\endcsname\relax\global\def\draftcontrol{0} 
\fi 

{\count255=\time\divide\count255 by 60 
\xdef\hourmin{\number\count255} 
\multiply\count255 by-60\advance\count255 by\time 
\xdef\hourmin{\hourmin:\ifnum\count255<10 0\fi\the\count255}} 
\def\draftdate{\number\month/\number\day/\number\year\ \ \ \hourmin } 

\newcommand\makepapertitle{\par

  \begingroup 
    \renewcommand\thefootnote{\@fnsymbol\c@footnote}%
    \def\@makefnmark{\rlap{\@textsuperscript{\normalfont\@thefnmark}}}%
    \long\def\@makefntext##1{\parindent 1em\noindent 
            \hb@xt@1.8em{%
                \hss\@textsuperscript{\normalfont\@thefnmark}}##1}%
     \newpage 
     \global\@topnum\z@   
     \@makepapertitle 
     \thispagestyle{empty}\@thanks 
  \endgroup 
  \setcounter{footnote}{0}%
  \global\let\thanks\relax 
  \global\let\makepapertitle\relax 
  \global\let\@makepapertitle\relax 
  \global\let\@thanks\@empty 
  \global\let\@author\@empty 
  \global\let\@date\@empty 
  \global\let\@title\@empty 
  \global\let\title\relax 
  \global\let\author\relax 
  \global\let\date\relax 
  \global\let\and\relax 
  \def\version{\let\version\@version\@gobble} 
} 
\def\@makepapertitle{%
  \newpage 
   \ifnum\draftcontrol=1 {} 
   \version\versionno 
   \vskip 5.5em%
   \else 
   \hfill\hbox to 3cm {\parbox{4.5cm}{\@pubnum}\hss}%
   \vskip 6.5em%
   \fi 
   \begin{center}%
   \let \footnote \thanks 
      {\hskip -0\textwidth \hbox to 1\textwidth%
        {\centerline{\Large\bf{\noindent\@title}}}}%
     \vskip 2em%
     {\normalsize
       \lineskip .5em%
       \begin{tabular}[t]{c}%
         \@author 
       \end{tabular}\par}%
     \vskip 1.5em%
     {\@bstract}%
     \end{center}%
     \vfill
     \@date%
     \vskip 1.5em%
   \par 
} 

\gdef\@pubnum{} 
\def\pubnum#1{%
  \gdef\@pubnum{#1}} 

\gdef\@bstract{} 
\def\Abstract#1{%
  \gdef\@bstract{%
   \parbox{\textwidth-0pc}{%
   \centerline{\bf Abstract}\penalty1000 
   \noindent
   \renewcommand\baselinestretch{1.0} 
   {#1}}} 
} 

\gdef\@email{}
\def\email#1{%
   \gdef\@email{%
   Email: {\tt #1}}
}

\def\ps@paper{\let\@mkboth\@gobbletwo%
     \ifnum\draftcontrol=1 
        \def\@oddfoot{\hbox to \textwidth{\tiny \versionno \hfil\tiny\draftdate}%
        \hskip -\textwidth \hbox to \textwidth{\hfil\rm\thepage\hfil}}%
     \else\def\@oddfoot{\hbox to \textwidth{\hfil\rm\thepage\hfil}} 
     \fi 
     \let\@evenfoot\@oddfoot 
} 

\def\body{\clearpage 
          \pagestyle{paper} 
        } 

\def\@version#1{\ifnum\draftcontrol=1 
\typeout{}\typeout{#1}\typeout{} 
\vskip3mm\centerline{\hbox{\fbox{\normalsize{\tt DRAFT -- #1 -- } 
                   {\draftdate}}}}\vskip3mm 
\fi} 
\let\version\@version 
\long\def\eqlabel#1{\ifnum\draftcontrol=1 
                    \tag@false  
                    \tag*{(\theequation) \hbox to -0.2cm{\hspace{0cm}\small{#1}\hss}} 
                    \refstepcounter{equation}  
                    \edef\@currentlabel{\theequation} 
                    \ltx@label{#1}          
                    \else 
                    \label{#1} 
                    \fi 
                    } 
\let\st@bibitem\@bibitem 
\let\st@lbibitem\@lbibitem 
\ifnum\draftcontrol=1 
  \def\@bibitem#1{%
    \st@bibitem{#1}\a@@label{#1}\ignorespaces} 
  \def\@lbibitem[#1]#2{%
    \st@lbibitem[#1]{#2}\a@@label{#2}\ignorespaces} 
  \def\a@@label#1{%
    \gdef\a@lab{\smash{\normalfont\small#1}} 
    \ifvmode 
      \if@inlabel 
        \global\setbox\@labels\hbox{%
          \llap{\a@lab\let\a@lab\relax 
                \kern\@totalleftmargin\kern\marginparsep}%
          \box\@labels}%
      \fi 
    \fi} 
\fi 

\documentclass[12pt,letterpaper]{article} 

\usepackage{amsmath,bm,amsfonts,amssymb,array,calc,amsthm,rotating,amscd,tikz-cd}
\usepackage{epsfig,psfrag} 
\usepackage{rotating}
\usepackage{amscd}
\usepackage{graphicx}
\usepackage{color}
\usepackage[colorlinks=false]{hyperref}

\renewcommand\baselinestretch{1.25} 
\renewcommand\section{\@startsection {section}{1}{\z@}%
                                   {-3.5ex \@plus -1ex \@minus -.2ex}%
                                   {2.3ex \@plus.2ex}%
                                   {\normalfont\large\bfseries}} 
\renewcommand\subsection{\@startsection{subsection}{2}{\z@}%
                                   {-3.25ex\@plus -1ex \@minus -.2ex}%
                                   {1.5ex \@plus .2ex}%
                                   {\normalfont\normalsize\bfseries}} 
\renewcommand\subsubsection{\@startsection{subsubsection}{3}{\z@}%
                                   {-3.25ex\@plus -1ex \@minus -.2ex}%
                                   {1.5ex \@plus .2ex}%
                                   {\normalfont\normalsize\it}} 
\renewcommand\paragraph{\@startsection{paragraph}{4}{\z@}%
                                   {-1.75ex\@plus -1ex \@minus -.2ex}%
                                   {1ex \@plus .2ex}%
                                   {\normalfont\normalsize\bf}} 
\renewcommand\subparagraph{\@startsection{subparagraph}{5}{\z@}%
                                   {-1.25ex\@plus -0ex \@minus -.2ex}%
                                   {-2ex \@plus .2ex}%
                                   {\normalfont\normalsize\it}}

\numberwithin{equation}{section}

\long\def\@makecaption#1#2{%
  \vskip\abovecaptionskip
  \sbox\@tempboxa{{\bf #1:} #2}%
  \ifdim \wd\@tempboxa >\hsize
    {\small\bf #1:} {\small #2}\par
  \else
    \global \@minipagefalse
    \hb@xt@\hsize{\hfil\box\@tempboxa\hfil}%
  \fi
  \vskip\belowcaptionskip}

\renewcommand*\l@section[2]{%
  \ifnum \c@tocdepth >\z@
    \addpenalty\@secpenalty
    \addvspace{.5em \@plus\p@}%
    \setlength\@tempdima{1.5em}%
    \begingroup
      \parindent \z@ \rightskip \@pnumwidth
      \parfillskip -\@pnumwidth
      \leavevmode \bfseries
      \advance\leftskip\@tempdima
      \hskip -\leftskip
      #1\nobreak\hfil \nobreak\hb@xt@\@pnumwidth{\hss #2}\par
    \endgroup
  \fi}
\renewcommand*\l@subsection{\addvspace{.0em \@plus\p@}\@dottedtocline{2}{1.5em}{2.3em}}
\renewcommand*\l@subsubsection{\addvspace{-.2em \@plus\p@}\@dottedtocline{3}{3.8em}{3.2em}}

\def\quantph#1{\href{http://xxx.arxiv.org/abs/quant-ph/#1}{{arXiv:quant-ph/#1}}}
\def\hepth#1{\href{http://xxx.arxiv.org/abs/hep-th/#1}{{arXiv:hep-th/#1}}}

\def\math#1{\href{http://xxx.arxiv.org/abs/math/#1}{{arXiv:math/#1}}}

\def\arxiv#1#2{\href{http://xxx.arxiv.org/abs/#1}{{arXiv:#1 [#2]}}}

\definecolor{refcol}{rgb}{0.2,0.2,0.8}
\definecolor{eqcol}{rgb}{.6,0,0}
\definecolor{purple}{cmyk}{0,1,0,0}

\gdef\@citecolor{refcol}
\gdef\@linkcolor{eqcol}
\def\colorlinkspurple{\gdef\@urlcolor{purple}}
\def\colorlinksblue{\gdef\@urlcolor{blue}}
\def\colorlinksred{\gdef\@urlcolor{red}}

\def\ie{{\it i.e.}}

\def\cf{{\it cf.}}

\def\revise#1       {\raisebox{-0em}{\rule{3pt}{1em}}%
                     \marginpar{\raisebox{.5em}{\vrule width3pt\ 
                     \vrule width0pt height 0pt depth0.5em 
                     \hbox to 0cm{\hspace{0cm}{%
                     \parbox[t]{4em}{\raggedright\footnotesize{#1}}}\hss}}}}

\def\ii           {{\it i}}

\def\Re           {{\rm Re\hskip0.1em}} 
\def\Im           {{\rm Im\hskip0.1em}}

\def\sqr#1#2{{\vcenter{\vbox{\hrule height.#2pt   
 \hbox{\vrule width.#2pt height#1pt \kern#1pt 
 \vrule width.#2pt}\hrule height.#2pt}}}}

\renewcommand{\P}{\mathbb P}
\newcommand{\R}{\mathbb R}
\newcommand{\Z}{\mathbb Z}

\newcommand{\C}{\mathbb C}

\newcommand{\N}{\mathbb N}

\newcommand{\Fcal}{\mathcal F}

\newcommand{\Ocal}{\mathcal O}
\newcommand{\Wcal}{\mathcal W}

\newcommand{\Ccal}{\mathcal C}
\newcommand{\Dcal}{\mathcal D}
\newcommand{\Ncal}{\mathcal N}

\newcommand{\Mcal}{\mathcal M}
\newcommand{\Ecal}{\mathcal E}

\newcommand{\ep}{\epsilon}

\newcommand{\beq}{\begin{equation}}
\newcommand{\eq}{\end{equation}}
\newcommand{\req}[1]{(\ref{#1})}

\newcommand{\bra}[1]{\left<#1\right|}
\newcommand{\ket}[1]{\left|#1\right>}
\newcommand{\cc}[1]{\overline{#1}}

\catcode`\@=12 

\begin{document} 


\title{Non-Perturbative Quantum Geometry III}

\pubnum{
CERN-TH-2016-100
}
\date{April 2016}

\author{
Daniel Krefl \\[0.2cm]
\it Theoretical Physics Department, CERN, Geneva, Switzerland \\
}

\Abstract{
The Nekrasov-Shatashvili limit of the refined topological string on toric Calabi-Yau manifolds and the resulting quantum geometry is studied from a non-perturbative perspective. The quantum differential and thus the quantum periods exhibit Stokes phenomena over the combined string coupling and quantized K\"ahler moduli space. We outline that the underlying formalism of exact quantization is generally applicable to points in moduli space featuring massless hypermultiplets, leading to non-perturbative band splitting. Our prime example is local $\P^1\times\P^1$ near a conifold point in moduli space. In particular, we will present numerical evidence that in a Stokes chamber of interest the string based quantum geometry reproduces the non-perturbative corrections for the Nekrasov-Shatashvili limit of 4d supersymmetric $SU(2)$ gauge theory at strong coupling found in the previous part of this series. A preliminary discussion of local $\P^2$ near the conifold point in moduli space is also provided.
}

\makepapertitle

\body

\version\versionno

\vskip 1em



\section{Introduction}
This work constitutes the third part of our series \cite{K13,K14} on the non-perturbative completion of the Nekrasov-Shatashvili (NS) limit of four dimensional gauge theories with eight supercharges in the $\Omega$-background, $\beta$-ensembles and refined topological strings. While we discussed mainly $\beta$-ensembles (at hand of the cubic) in the first part \cite{K13}, gauge theories (using $SU(2)$ as illustrative example) in the second part, in this work we move on to refined topological strings on toric Calabi-Yau manifolds.

The unifying theme of this series is the underlying perturbative (semi-classical) quantum geometry (in the sense of \cite{MM09,ACDKV11}), completely describing the NS limit of the above theories. The models discussed in \cite{K13,K14} are on the level of the quantum geometry essentially equivalent to well-known and simple one dimensional quantum mechanical systems. Therefore the by now extensive knowledge about exact quantization of such systems (see \cite{ZJ04a,ZJ04b} and references therein) was harvested in \cite{K13,K14} to infer the non-perturbative completion of the models under consideration, which lead to the notion of non-perturbative quantum geometry. Another notable work about the $SU(2)$ case along similar lines is \cite{BD15}.

On a more technical level, the cubic $\beta$-ensemble and $SU(2)$ gauge theory have two properties in common. Firstly, both models feature two underlying moduli, out of which one is analytically continued to negative values. In case of the $\beta$-ensemble the moduli are the number of eigenvalues localized in each of the two cuts, and for $SU(2)$ the two Cartans (vector multiplet moduli $a_i$ with $a_1=-a_2$). Secondly, both models feature massless hyper- or vector-multiplets at the point of expansion considered (in the case of $\beta$-ensembles these correspond to the gaussian normalization factors). As we will make more explicit in this work, the crucial properties for the non-perturbative completion are the massless multiplet contributions to the quantum B-periods and the analytic continuation thereof. This tells us that the same approach should be viable to toric Calabi-Yaus, at least at particular points in moduli space featuring massless states (as well as for specific phases of the moduli and $\hbar$ in the complex plane). The prime example is the conifold point in moduli space. 

In detail, the massless multiplet contribution in the NS limit possesses a $\log \Gamma$-function term depending on the moduli combination $t/\hbar$, where $t$ is the flat coordinate near the point of expansion. Under the analytic continuation $t/\hbar\rightarrow -t/\hbar$ the quantum B-period picks up an infinite series of non-perturbative corrections in powers of $\zeta=e^{-\frac{\ii\pi t}{\hbar}}$, due to Eulers reflection formula for the $\Gamma$-function. One might see these corrections as an NS analog of the so-called A-cycle instanton contributions, which have been extensively discussed for topological strings and matrix models in \cite{PS09}. 

The perturbative quantum geometry is based on the WKB solution of the wave equation arising from quantizing the underlying geometry (algebraic curve). It was observed in \cite{K13,K14} that (at least for the models under consideration) the exact quantization condition, which essentially leads to normalizable bound or resonance state wave-function solutions, is equivalent to the NS quantization condition \cite{NS09}
\beq\eqlabel{NScondOrig}
\partial_{t_i} \Wcal(t,\hbar):=\frac{\Pi_{B_i}(t,\hbar)}{\hbar} = 2\pi \ii \, n_i\,,
\eq
where $\Pi_B$ denotes a quantum B-period, $n_i\in\Z$ and $\Wcal$ is the NS free energy. One may also see the NS condition simply as the requirement that the wave-function does not pick up a phase under monodromy along a B-cycle.

The NS (exact quantization) condition constrains the moduli of the system, as the $t_i$ have to take particular values such that \req{NScondOrig} is satisfied. To take an example that has been extensively discussed in the previous parts of this series: if only a single effective modulus is present, the corresponding quantum A-period becomes quantized (the usual Bohr-Sommerfeld condition), and also receives non-perturbative corrections in powers of $\xi = e^{-\frac{c_X}{\hbar}}$ (where $c_X$ is some constant), \ie,  $t\sim \hbar N +\Ocal(\xi)$ ($N$ is an integer), at a point in moduli space with massless hypermultiplets present. Inserting the corrected $t$ (which is the non-perturbatively flat coordinate) into the dual B-period, we infer that
$$
\zeta \rightarrow (-1)^N(1+\Ocal(\xi)) \,.
$$
Hence, exact quantization ensures that both the A- and B-period are well-defined trans-series in $\xi$.

There is one important point which has not been explicitly mentioned in the previous parts of this series, namely that there is a sign degree of freedom in exact quantization, \ie, the wave-function may also behave anti-periodic under monodromy. In particular, this leads to the band splitting of energy levels $E\rightarrow E^\pm$, where the levels differ by the non-perturbative terms in powers of $\xi$. As the NS quantization condition only dictates the presence of one of the two bands (integer quantization), one might ask if the other band (half-integer quantization) is a physically viable solution, or merely a mathematical curiosity. We leave the answer to this question to follow-up works and in the rest of this paper we will consider both signs in order to be fully general.  

The main purpose of the present work is to give some deeper insights into the non-perturbative quantum geometry introduced in \cite{ACDKV11,K13,K14}, which encodes the NS limit of physical theories. In particular, we find it important to stress that the quantum curves are actually complex. One implication is that taking a single real slice of the geometry is usually not sufficient to obtain a complete picture. This is already very clear in the case of the deformed conifold. Depending on how we take the real slice, and at what point we sit in the combined $\hbar$ and complex structure moduli space, we either obtain the 1d quantum theory of the harmonic oscillator or the parabolic barrier, which are fundamentally different. In reality, the complex quantum theory interpolates between both. This means that the wave-function solutions, and thus the quantum differential, will have phase transitions between a bound state and a resonance solution, depending on where we sit in the combined moduli space. In particular, the presence of non-trivial non-perturbative effects depends on which solution we consider. This insight is key to understanding how to reproduce the non-perturbative completion of $SU(2)$ gauge theory developed in \cite{K14} via geometric engineering. Furthermore, normalizability of the wave-function shows a richer solution set in the complex setting, since we have a choice of path and singularities to connect (the path between two singularities on which to normalize the wave-function). In this work we will only give a rough but solid sketch of the underlying fundamental story, leaving the task to working out a detailed formalisation to future research.

The outline of this paper is as follows: In the next section we will briefly review the formalism of quantum geometry along the lines of \cite{ACDKV11,K14}, with special emphasis on the occuring Stokes phenomena as the quantum modulus (to be defined in section \ref{QGsec}) and the coupling constant $\hbar$ are varied. In section \ref{DefConiSec} we will discuss the square potential (geometrically corresponding to the deformed conifold), which is the core example from which the essential aspects of the non-perturbative completion can already be inferred. In particular, we will explain in this section why the formalism of \cite{K13,K14} extends to toric Calabi-Yaus featuring a conifold singularity in their B-model complex structure moduli space and beyond. In section \ref{flToricCYsec} we will take a first detailed look at a toric Calabi-Yau, namely local $\P^1\times\P^1$. It is well known that this geometry engineers a pure $SU(2)$ gauge theory \cite{KKV96}. Therefore, one would expect to be able to reproduce the non-perturbative corrections found for $SU(2)$ in \cite{K14} from this string geometry. We will show numerically that this is indeed the case, by chosing a suitable wave-function basis. Furthermore, we will present evidence that  even away from the gauge theory limit non-perturbative effects are present and lead to band splitting for this wave-function basis. In contrast, the other possible basis (harmonic oscillator expansion) is not corrected but is instead calculable order by order via WKB. The final section, section \ref{P2sec}, is used to present some preliminary results for local $\P^2$. In appendix \ref{NumericSec} some more details about the numerical techniques invoked in this work are given.

\section{Quantum Geometry}
\label{QGsec}

Consider an algebraic curve
$$
\Sigma: f(x,p)=0\,,
$$
in $\C^*\times\C^*$, not necessarily polynomial. We should think about the curve as being a fibration over the complex structure moduli space $\Mcal$, \ie,
$$
\Sigma\rightarrow \Mcal\,.
$$
We can always arrange via appropriately transforming the curve that one modulus is separated. We will refer to this modulus as the quantum modulus, denoted in the following as $E$. The perhaps simplest example is the curve
\beq\eqlabel{GaussianCurve}
p^2+x^2=E\,,
\eq
with the quantum modulus $E\in\Mcal = \C$. 

Canonical quantization of $\Sigma$ amounts to promoting the coordinates $x$ and $p$ to anti-commuting operators $[x,p]\neq 0$. In general we take
$$
[x,p]=\ii\hbar=\ii|\hbar|e^{\ii\theta}\in\C\,.
$$ 
Hence, $p\sim \partial_x$ and the curve turns into a differential (or, if exponentials are involed, difference) operator $\Dcal_\hbar$ eigenvalue problem with solutions $\Psi^{(i)}(x)$ and eigenvalues given by the quantum modulus. 
Note that in this work we only consider curves which yield under quantization an operator of second order.

Differentials $dS$ on $\Sigma$ can be defined via 
\beq\eqlabel{dSdef}
dS \sim \partial_x\log \Psi(x)\,,
\eq
where $\Psi$ denotes a particular linear combination of the solutions (wave-functions), \ie, $\Psi(x):=\sum_{i}c_i\Psi^{(i)}(x)$. Possible linear combinations are constraint by the requirement that $\Psi$ decays fast enough at infinity. However, as we are in the complex setting, a richer structure of (normalizable) solutions emerges than in ordinary quantum mechanics. This is because in requiring 
\beq\eqlabel{squareintCond}
\int_{\Ccal}dx\,\cc{\Psi(x)}\Psi(x)<\infty\,,
\eq
we have the freedom to tune to a suitable contour $\Ccal$ connecting different infinities of $\Sigma$, instead of being constraint to the real line. In particular, in general there will be multiple solutions connecting different pairings of inequvialent singularities.

A quantum curve is defined as the classical curve $\Sigma$ equipped with one of the differentials, \ie, a pair $(\Sigma, dS)$. Periods $\Pi$ can then be obtained as usual via integrating over closed cycles,
$$
\Pi = \oint dS\,.
$$
Note that consistent solutions $\Psi$ and so differentials $dS$ may not exist for all points in $\Mcal$. Therefore the moduli space $\Mcal_\hbar$ of the quantum curve $(\Sigma, dS)$ is in general only a, perhaps discrete, subspace of $\Mcal$, \ie, $\Mcal_\hbar\subset \Mcal$. More specifically, usually consistent $\Psi$ require a discrete and $\hbar$ dependent quantum modulus. One should view, both, the quantum curve and the reduced moduli space $\Mcal_\hbar$ as a fibration over $\C$, the value $\hbar$ the commutator takes. Pictorially, 
\beq\eqlabel{CopSquarePot}
\begin{tikzcd}
&(\Sigma, dS)\arrow{rd}{} \arrow{r}{}& \Mcal_\hbar\arrow{d}\\
&&\C
\end{tikzcd} 
\,.
\eq
Generally, the quantum curve $(\Sigma, dS)$, or more specifically the differential $dS$, is not smooth,  both, under variations of the quantum modulus and of $\hbar$, \ie, exhibits phase transitions (Stokes phenomena). The reason is that the underlying consistent linear combination of solutions may jump under varying the moduli (including $\hbar$). We will illustrate this fact in detail at hand of an explicit example in section \ref{DefConiSec}. 

In general, we do not know the wave-functions and so the quantum differentials exactly, but rather only a WKB approximation for $|\hbar|$ small of the solutions of $\Dcal_\hbar$, \ie,
\beq
\eqlabel{WKBansatz}
\Psi^\pm_{\rm WKB}(x) \sim e^{\pm\frac{1}{\hbar}\int^x dS}\,,
\eq
such that $\Psi\sim c_+ \Psi^+_{\rm WKB}+c_- \Psi_{\rm WKB}^-$. Hence, the differential is expanded for small $|\hbar|$,
$$
dS=\sum_{n=0}^\infty dS^{(n)}\, \hbar^n\,,
$$ 
and so the periods, leading to a semi-classical approximation, which is usually an asymptotic expansion. Note that the rationale for the definition of the differential $dS$ \req{dSdef} can be seen as rooted in the WKB Ansatz for $\Psi$.

The WKB Ansatz introduces an additional complication, as in the WKB approximation the linear combination corresponding to a consistent $\Psi$ may not only jump under varying $E$ and $\hbar$, but generally as fibration over $\Sigma$. However, one can infact use this property to derive a condition on the moduli, as a consistent solution requires that under analytic continuation over $\Sigma$ a wave-function decaying at one infinity in the complex plane continues to a decaying wave-function at another infinity. Essentially, this is what exact quantization is about, \ie, finding bound states or resonances. 

Surprisingly, as first found and used in \cite{K13}, for certain models of interest with one effective modulus, the exact quantiziation condition is equivalent to the Nekrasov-Shatashvili quantization condition (\req{NScondOrig} exponentiated) \cite{NS09}
\beq\eqlabel{NScond}
\phi:= e^{\frac{\Pi_{B}(E)}{\hbar}}= 1\,.
\eq
Physically, the NS condition ensures that we sit in a supersymmetric vacuum of the corresponding effective 2d theory. It was further proposed in \cite{K13} to use this condition in general, including for toric Calabi-Yaus, to infer non-perturbative information, simply because \req{NScond} can only be satisfied if the quantum modulus $E$ receives non-perturbative corrections. Note that it is clear from \req{WKBansatz} that $\phi$ corresponds to the phase of the WKB wave-function under monodromy along the B-cycle.

In exact WKB \req{NScond} is however not the unique condition for the existence of bound states/resonances, as $\phi=-1$ is another possibility. At the time being, we do not understand the meaning of $\phi=-1$ in the effective 2d theory, \ie, if half-integer values for the derivative of the effective superpotential (that is $n_i \in \Z/2$ in \req{NScondOrig}) form consistent solutions which have been overlooked in the literature or not (\cf, \cite{NS09a}). However, we know that mathematically both boundary conditions are consistent (\cf, \cite{DDP97} and references therein), and in particular lead to energy band splitting,
$$
E\rightarrow E^{\pm}\,.
$$
This can be verify numerically, for instance for the Mathieu equation and thus $SU(2)$ gauge theory in the NS limit. Therefore, we impose in general as quantization condition $\phi = \pm 1$ in the following sections.

Note that in general we do not know the exact $\Psi$ (or $\Psi_{\rm WKB}$) as a function of the complex structure moduli, but rather expansions thereof at particular points in the moduli space $\Mcal_\hbar$. The different expansions can be obtained via reparameterizations of the curve $\Sigma$ (as we will illustrate for some examples below) and hence the WKB expansions at different points are related by the underlying modular group of the curve. However, we like to stress here that the physical nature of the non-perturbative effects, ensuring that \req{ExactQuantCond} holds, change over moduli space, see \cite{K14} and in particular \cite{BD15}.

\section{Deformed conifold}
\label{DefConiSec}
\paragraph{Exact square potential}
Consider the curve \req{GaussianCurve}. It is instructive to introduce an additional parameter $\omega$ and take as curve $p^2+\omega^2 x^2=E$. Quantization gives the operator 
$$
\Dcal_\hbar: \frac{\partial^2}{\partial x^2} -\kappa^2 x^2 +\frac{E}{\hbar^2}   \,,
$$
where we defined for convenience the parameter
$$
\kappa := \frac{\omega}{\hbar}\,.
$$
The above operator leads to Weber's equation with the two independent solutions given in terms of the parabolic cylinder functions $D(N,x)$ by
\beq
\begin{split}
\Psi^{+}(x) &= D\left(\frac{1}{2}\left(\frac{E}{\hbar\omega}-1\right) , \sqrt{2\kappa}\, x\right)\,,\\
\Psi^{-}(x) &= D\left(-\frac{1}{2}\left(\frac{E}{\hbar\omega}+1\right) , \ii  \sqrt{2\kappa}\, x\right)\,.
\end{split}
\eq

In general $\Psi = c_+ \Psi^{+}+c_-\Psi^{-}$ is not square-integrable for arbitrary $\frac{E}{\hbar\omega}\in\C$ and for arbitrary integration contours connecting infinities in the complex plane. However, square integrable solutions can be found as follows. For $N\in\N$ the cylinder functions reduce to the Hermite polynomials $H_N(x)$, \ie, 
$$
D(N,x) = \frac{1}{\sqrt{2^N}}\,e^{-\frac{x^2}{4}}\, H_N\left(\frac{x}{\sqrt{2}}\right)\,,
$$
which satisfy $H_N(-x)=(-1)^N H_N(x)$ and the orthogonality relation
$$
\int_{-\infty}^\infty dx\, e^{-x^2}H_m(x) H_n(x) = 2^n n!\sqrt{\pi} \, \delta_{nm}\,.
$$
In particular we have $\int_{-\infty}^\infty dx\,  e^{-x^2} |H_m(x)|^2 <\infty$. Hence, for $\hbar\omega\in\R^+$ such that $E=\hbar\omega(2N+1) >0$ with $N\in \N$ we have a square integrable solution (over the real line) $\Psi_{E>0}\sim \Psi^{+}$, with associated differential $dS_{E>0}$ fibered over a discrete subspace $\Mcal_\hbar$ of $\Mcal$. The normalization can be fixed via the above orthogonality relation, yielding
\beq\eqlabel{HOwfEp}
\Psi_{E>0}(x)=\frac{1}{\sqrt{2^{N}N!}}\left(\frac{\kappa}{\pi}\right)^{1/4} e^{-\frac{\kappa x^2}{2}}H_N\left(\sqrt{\kappa} x\right)\,.
\eq
These are just the usual bound state solutions of the harmonic oscillator.

Additionally, we have another solution set with $\hbar\omega\in\R^+$ in terms of the Hermite polynomials for negative real energies $E=-\hbar\omega(2N+1)<0$, where $\Psi_{E<0}\sim \Psi^{-}$ is the wave-function. The normalized solution reads
$$
\Psi_{E<0}=\frac{1}{\ii^N\sqrt{2^{N}N!}}\left(\frac{\kappa}{\pi}\right)^{1/4} e^{\frac{\kappa x^2}{2}}H_N\left(\ii \sqrt{\kappa} x\right)\,.
$$
As these solutions are not square-integrable on the real line, they are usually discarded as unphysical in ordinary quantum mechanics. As we are here in the complex setting, we are less constraint. For instance, $\Psi_{E<0}$ is square integrable instead on the imaginary axis. More generally, in the Stokes chamber with $|\Im x| > |\Re x|$. 

We conclude that there is a phase transition (Stokes phenomena) under analytically continuing the modulus $E$ through zero. That is, the quantum differential and so the quantum periods jump. Up to normalization we can also rotate the solutions via $\kappa$ (and so $\hbar$) into each other, \ie,
\beq
\begin{tikzcd}
\Psi_{E<0}\arrow[leftrightarrow]{r}{\kappa\rightarrow -\kappa}& \Psi_{E>0}\\
\end{tikzcd} 
\,.
\eq

So far, we considered $\hbar w$ and so $\kappa$ to be real. However, we can also rotate $\kappa$ into the complex plane. For instance, rotating $\kappa\rightarrow \ii\kappa$ we obtain the solutions of the inverted harmonic oscillator (parabolic barrier), which we will denote as $\Psi^*$ and which have imaginary energy. Indeed, comparing with the solutions given for instance in \cite{BF13}, we infer that up to normalization
\beq
\begin{tikzcd}
\Psi_{E<0}\arrow[leftrightarrow]{r}{\kappa\rightarrow -\kappa}\arrow[leftrightarrow, swap]{d}{\kappa\rightarrow\ii\kappa}& \Psi_{E>0}\arrow[leftrightarrow]{d}{\kappa\rightarrow\ii\kappa}\\
\Psi^*_{\ii E>0}\arrow[leftrightarrow, swap]{r}{\kappa\rightarrow -\kappa}&\Psi^*_{\ii E<0}\\
\end{tikzcd} 
\,.
\eq
Again, in the complex plane these solutions turn normalizable, as we can adjust the contour accordingly. More generally, for arbitrary complex $\kappa$ the two fundamental solutions $\Psi^{+}$ and $\Psi^{-}$ are normalizable on the (discrete) line $E=\pm \hbar^2\kappa(2N+1)\subset\Mcal$. In particular, we have that
$$
\Psi^- = \Psi^+(\kappa\rightarrow -\kappa)\,.
$$
The integration contours for \req{squareintCond} depends on the value $\kappa$ takes. Fixing $|\Re\kappa|>|\Im\kappa|$, the solution $\Psi^+$ can be integrated over $\Ccal_+=[-\infty,\infty]$ while $\Psi^-$ over $\Ccal_-=[-\ii\infty,\ii\infty]$. Under varying the modulus $\kappa$ and/or $E$ we encounter Stokes phenomena switching between $\Psi^{\pm}$. 

We conclude that in this example the quantum differential $dS$ has indeed a non-trivial phase structure over the combined moduli space of $\hbar$ and $E$.

\paragraph{WKB square potential}

Let us now consider the same curve \req{GaussianCurve} in the WKB approximation for the quantum differential. It is convenient to go to a point in the $\hbar$ moduli space such that we have the quantization condition $[x,p]=-\hbar$ with $\hbar\in\N$. The leading order of the differential $dS$ can be easily inferred from the WKB Ansatz \req{WKBansatz} to be 
$$
dS^{(0)} = \ii\sqrt{E-x^2}\,.
$$
The curve has branch points at $x=\pm \sqrt{E}$ and therefore we have an A-period 
$$
\Pi_A = \frac{1}{\pi\ii}\int_{-\sqrt{E}}^{\sqrt{E}}dS = \frac{E}{2} +\Ocal(\hbar)\,,
$$
and a dual B-period
$$
\Pi_B = 2\int_{\sqrt{E}}^{\Lambda}dS = \frac{E}{2}\left(1-\log\left(\frac{E}{4 \Lambda^2}\right) \right)  -\Lambda^2 +\Ocal(\hbar)\,,
$$
where we introduced a cutoff $\Lambda$, which will play an important role. In detail, keeping $\Lambda$ finite simulates general geometries, as we can approximate the potential barrier between two vacua by a finite inverse square potential.

The higher order corrections to the periods can be inferred for instance as in \cite{ACDKV11} via deriving operators $\Dcal^{(n)}$ with $\Pi^{(n)} = \Dcal^{(n)}\Pi^{(0)}$. We just state the result here. $\Pi_A$ as given above is perturbatively exact, while the first few orders of $\Pi_B$ read
$$
\Pi_B=-\Lambda^2+\frac{E}{2}\left(1-\log\left(\frac{E}{4 \Lambda^2}\right) \right) -\frac{1}{12} \frac{\hbar^2}{E}-\frac{7}{360}\frac{\hbar^4}{E^3}-\frac{31}{1260}\frac{\hbar^6}{E^5} +\Ocal(\hbar^8)\,.
$$
Note that the cutoff $\Lambda$ only enters at order $\hbar^0$. Comparing with the asymptotic expansion of $\log\Gamma$ for large arguments, we infer that infact $\Pi_B$ corresponds thereto, \ie,
\beq\eqlabel{GaussianBperiod}
\frac{1}{\ii \hbar}\Pi_B(E) = \log\Gamma\left(\frac{1}{2}+\frac{\ii E}{2\hbar}\right) +\frac{E}{\hbar}\log\left(\frac{\Lambda}{\hbar}\right) -\frac{\Lambda^2}{\ii\hbar}-\frac{1}{2}\log 2\pi\,,
\eq
is the exact B-period. 

The phase transition discussed above follows under varying $\frac{\ii E}{\hbar}$ into the negative domain of the complex plane from Euler's reflection formula for the $\Gamma$-function
\beq\eqlabel{EulerRefl}
\Gamma(z)\Gamma(1-z)=\frac{\pi}{\sin(\pi z)}\,.
\eq
For instance, under $E\rightarrow -E$ we have
\beq
\begin{split}
\Pi_B(-E) &= \Pi_B(E) -2\ii E\log\left(\frac{\Lambda}{\hbar}\right) + \ii\hbar \log \cos\left(\frac{\pi \ii E}{2\hbar}\right) -\ii\hbar\log\pi  \\
&=\Pi_B(E) -2\ii E\log\left(\frac{\Lambda}{\hbar}\right)+ \frac{\pi E}{2}-\ii\hbar \sum_{k=1}^\infty \frac{e^{-\frac{k \pi E}{\hbar}}}{k} -\ii\hbar\log 2\pi \,.
\end{split}
\eq
We observe that the $B$-period picks up an infinite series of non-perturbative corrections in powers of $\zeta:=e^{-\frac{\pi E}{\hbar}}$ under the analytic continuation. Note that these terms are independent of $\Lambda$. 

The NS quantization condition \req{NScondOrig} can be satisfied for $\Pi_B(-E)$. In detail, following the previous parts of this series \cite{K13,K14}, for $E<0$ we impose the exact quantization condition (uniqueness of the wave-function) 
\beq\eqlabel{ExactQuantCond}
e^{\frac{\Pi_B(-\Ecal)}{\ii\hbar}} = \pm 1\,,
\eq
where $\Ecal$ denotes a new coordinate, which is non-perturbatively flat, \ie,
$$
\Ecal = E + E_{np}\,.
$$
$E_{np}$ denote the non-perturbative corrections to the perturbatively flat coordinate $E$. Inserting \req{GaussianBperiod} into \req{ExactQuantCond}, defining an instanton counting parameter
\beq\eqlabel{GaussianNPpara}
\xi:=e^{-\frac{\Lambda^2}{\ii\hbar}}\,,
\eq
the exact quantization condition turns into
$$
\frac{\cos\left(\frac{\pi \ii \Ecal}{2\hbar}\right)}{\pi}= \pm \frac{1}{\sqrt{2\pi}\, \Gamma\left(\frac{1}{2}+\frac{\ii \Ecal}{2\hbar}\right)}\, \left(\frac{\Lambda}{\hbar}\right)^{-\Ecal/\hbar} \, \xi\,,
$$
where we made use of Euler's reflection formula. The above relation can be solved via expanding $E_{np}$ into powers of $\xi$ such that
$$
\Ecal^\pm = E + \sum_{n=1}^\infty E^{(n)}_{\pm, np}\,\xi^n\,.
$$
The non-perturbative contributions $E^{(n)}_{\pm, np}$ then can be obtained order by order in $\xi$, as in \cite{K13,K14}. For instance, at order $\xi^0$ we recover the usual Bohr-Sommerfeld quantization condition
$$
\Pi_A= \frac{E}{2}=\ii\hbar(2N+1)+\Ocal(\xi^1)\,.
$$
Note that for $\Lambda\rightarrow\infty$ we have $\xi\rightarrow 0$ and hence the corrections to the A-period vanish, \ie, $\Ecal = E$. This is in accord with our exact discussion about the square potential above.

Finally, we evaluate $\Pi_B$ at the non-perturbatively flat coordinate $-\Ecal$. We see that in fact $\Pi_B$ turns into a trans-series in terms of the original coordinate $E$, \ie,
$$
\Pi_B(-\Ecal)=\Pi_B(E+\Ocal(\xi)) - 2\ii (E+\Ocal(\xi)) \log\left(\frac{\Lambda}{\hbar}\right)-\ii\hbar\log\pi -\hbar\pi +\Ocal(\xi)\,.
$$ 
Hence, the exact quantization leads to a well-defined trans-series expansion of, both, the A- and B-period in terms of $\xi$.

\paragraph{Toric Calabi-Yaus at a conifold point}
It is well known that the free energy of the topological string on a deformed conifold can be obtained from a Schwinger integral due to integrating out a hypermultiplet which becomes massless \cite{V95,GV95}. Similarly, the contribution of a single massless hypermultiplet to the free energy in the $\Omega$-background, and thus the refined topological string on a deformed conifold, is governed by the Schwinger integral (see for instance \cite{KW10b})
\beq\eqlabel{Fhyper}
\Fcal_{sing}(t) = \frac{1}{4}\int_\delta^{\infty} \frac{dx}{x} \frac{e^{-t x}}{\sinh\left(\frac{\ep_1 x}{2}\right)\sinh\left(\frac{\ep_2 x}{2}\right)}\,,
\eq
where $\delta \rightarrow 0$ is a cutoff. The NS limit of the corresponding B-period is easily taken, yielding
$$
\mathcal \partial_t W_{sing}(t) :=\lim_{\ep_2\rightarrow 0}\ep_2\, \partial_t\Fcal(t) = -\frac{1}{2}\int_\delta^\infty \frac{dx}{x}\frac{e^{-t x}}{\sinh\left(\frac{\hbar x}{2}\right)}= -\int_\delta^\infty \frac{dx}{x}\frac{e^{-(t-\hbar/2 )x}}{e^{\hbar x} -1}\,,
$$
where we set $\hbar:=\ep_1$. Integrals of this kind have been discussed in the NS context before in \cite{KS13,K13}, from which we can infer that
$$
\partial_t\Wcal_{sing}(t) = -\frac{\hbar}{\delta}- \frac{t}{\hbar}\left(\log\frac{\delta}{\hbar}+\gamma\right)-\log \Gamma\left(\frac{1}{2}+\frac{t}{\hbar}\right) +\frac{1}{2}\log 2\pi\,,
$$
where $\gamma$ denotes the Euler-Mascheroni constant. The singular terms for $\delta\rightarrow 0$ are an artifact of the fact that the Schwinger integral \req{Fhyper} commonly used in the literature is not properly regularized. Therefore we simply drop these singular terms such that we arrive at 
$$
\partial_t\Wcal_{sing}(t) = -\frac{t}{\hbar}\left(\log\frac{1}{\hbar}+\gamma\right)-\log \Gamma\left(\frac{1}{2}+\frac{t}{\hbar}\right) +\frac{1}{2}\log 2\pi\,.
$$
Note that this result can also be derived using the better behaved expression for the multiplet contribution derived in the context of gauge theory in \cite{NO03,NY03}, which yield precisely the above finite expression (see appendix C of \cite{K14}), up to the term proportional to $\gamma$. We suspect that this term is another artifact of the improper regularization of \req{Fhyper}, and therefore drop it as well. Taking the exponential yields
$$
e^{-\partial_t\Wcal_{sing}(t)} = \frac{1}{\sqrt{2\pi}}\left(\frac{1}{\hbar}\right)^{\frac{t}{\hbar}}\Gamma\left(\frac{1}{2}+\frac{t}{\hbar}\right) \,.
$$
The above result is important, because the B-model complex structure moduli space of toric Calabi-Yaus possess in general conifold points, where a deformed conifold singularity emerges. The singular terms of the refined topological string expanded near such points in moduli space are precisely captured by the Schwinger integral of integrating out a massless hypermultiplet in the $\Omega$-background given in \req{Fhyper}. However, additional regular terms will be present due to the embedding into the Calabi-Yau, \ie, in general we have
$$
\partial_t\Wcal(t) = \partial_t\Wcal_{sing}(t)+\partial_t\Wcal_{reg}(t)\,,
$$
where essentially $\partial_t\Wcal = \frac{\Pi_B}{\hbar}$ (\cf, \cite{ACDKV11}). Usually, the regular terms at the conifold point in moduli space go like
$$
\partial_t\Wcal_{reg}(t) = \frac{c_X}{\hbar} + A_p(t)\,,
$$
where $c_X$ refers to the leading non-singular term, which is generally a constant at the conifold point in moduli space and $A_p$ refers to the remaining regular contributions.

Hence, from the previous discussions we immediately infer that
$$
e^{-\partial_t\Wcal(t)} = \frac{1}{\sqrt{2\pi}}\left(\frac{1}{\hbar}\right)^{\frac{t}{\hbar}}\Gamma\left(\frac{1}{2}+\frac{t}{\hbar}\right) e^{-A_p(t)}\, \xi \,,
$$
where we defined as instanton counting parameter
$$
\xi:= e^{-\frac{c_X}{\hbar}}\,.
$$
As the embedding of the conifold singularity into a Calabi-Yau provides a cutoff, this result is equivalent to the pure deformed conifold discussed earlier, with cutoff $\Lambda^2=\partial_t W_{reg}$. Under the analytic continuation $t/\hbar\rightarrow -t/\hbar$ the free energy $\Wcal(t)$ acquires non-perturbative corrections due to Euler's reflection formula \req{EulerRefl}, \ie,
\beq\eqlabel{Wtransf}
\partial_t\Wcal_{sing}(-t/\hbar)=\partial_t\Wcal_{sing}(t/\hbar)+\partial_t\Wcal_{np}(t/\hbar)\,.
\eq
We can impose the exact quantization condition, and calculate as in the previous parts of this series non-perturbative corrections to the flat coordinate $t$ at the conifold point in moduli space, order by order in $\xi$. 

\paragraph{Beyond conifold points}
It is clear that the formalism extends to other points in moduli space and perhaps even beyond the NS limit. The technical details and physical nature of the non-perturbative effects will differ to some extend, see in particular \cite{BD15}. However, the key underlying concepts, namely the analytic continuation of the moduli in the complex plane, the occuring phase transitions under which the B-period picks up non-perturbative corrections and the need to introduce a non-perturbatively flat coordinate does not change. 

It is instructive to consider again the Schwinger integral \req{Fhyper}. Under $\hbar\rightarrow \ii\hbar$ we have that 
$$
\sinh\rightarrow \ii \sin \,,
$$
and thereby obtaining poles on the integration axis. Similar as in Schwinger's original work, deforming the contour to avoid the poles we pick up in the asymptotic expansion a non-perturbative contribution
\beq
\begin{split}
\partial_t\Wcal_{np}(t) &=\frac{\pi}{4}{\rm Res}_{x= \frac{2 \pi n}{\hbar}}\, \frac{e^{-t x}}{x\sin\left(\frac{\hbar x}{2}\right)}  + \frac{\pi}{2}\sum_{n=1}^\infty {\rm Res}_{x= \frac{2 \pi n}{\hbar}}\, \frac{e^{-t x}}{x\sin\left(\frac{\hbar x}{2}\right)}\,.
\end{split}
\eq
Because $\delta\rightarrow 0$ we also take along the pole at zero, however, with an additional factor of $1/2$ as we should only correct by a quarter circle at zero. Calculating the residue yields
$$
\partial_t\Wcal_{np}(t) =-\frac{t \pi}{2h} +\frac{1}{2}\sum_{n=1}^\infty \frac{(-1)^n}{n}e^{-\frac{2\pi n t}{\hbar}} = -\frac{1}{2}\log \cos\left(\frac{\pi t}{\hbar}\right) -\frac{1}{2}\log 2\,.
$$
Up to some constant this is precisely the non-perturbative term of the $\log\Gamma$-function on the Stokes line, \cf, \cite{PW91,N13}. The underlying important relation is the transformation \req{Wtransf} under varying the moduli $t$ and $\hbar$. Since a slight modification of the Schwinger integral captures the (refined) topological BPS expansion at the large volume point in moduli space, in general we will pick up an additional $\Wcal_{np}$ under the analytic continuation (more precisely from the tree-level part, as discussed extensively for example in \cite{K15b}). Enforcing the NS (or exact) quantization condition \req{NScondOrig}, then leads to a non-perturbatively corrected flat coordinate and so to a trans-series expansion of both the A- and B-period. This is the story more or less already envisaged in our first part \cite{K13} of the current series. However, we leave the general details at the large volume point in moduli space still to future work, and instead consider here toric Calabi-Yau examples at the conifold point in some more detail, thereby learning important lessons.

\section{A first look at a toric Calabi-Yau}
\label{flToricCYsec}

Let us consider the classical curve
\beq\eqlabel{F0origCurve}
\Sigma: -1+e^x+e^p+z_1e^{-x}+z_2e^{-p}=0\,.
\eq
This geometry corresponds to the mirror curve of local $\P^1\times \P^1$. The parameterization used is convenient, as we can directly extract via expansion for small $z_i$ the perturbative (quantum) periods at the large volume point in moduli space, \cf, \cite{ACDKV11}.

Let us however change parameterization of the curve as follows. We redefine
$$
x\rightarrow \ii x+\frac{1}{2}\log z_1\,,\,\,\,\, p\rightarrow p+\frac{1}{2}\log{z_2}\,.
$$
The curve turns into
\beq\eqlabel{clCurve}
2\lambda\cos(x)+ e^p+e^{-p}=E\,,
\eq
where we defined $\lambda:=\ii \sqrt{\frac{z_1}{z_2}}$ and $E:=\frac{1}{\sqrt{z_2}}$. Note that at $z_1=z_2$ we have $\lambda=\ii$ and recover (up to the reparameterization $x\rightarrow \ii x$) the curve used for instance extensively in \cite{HW14}. The large volume regime corresponds to $E\gg 1$ with $\lambda\sim \ii$. For us, it will be important to keep explicitly $\lambda$ as a free parameter. The quantization,
\beq\eqlabel{QuantCond}
[x,p]=\ii\hbar\,,
\eq
amounts to lift to a difference operator
$$
e^p+e^{-p}\rightarrow \Dcal = e^{\ii\hbar\partial_x}+e^{-\ii\hbar\partial_x} \,,
$$
such that we obtain the eigenvalue problem
\beq\eqlabel{qMathieuEq}
\left(\Dcal+2 \lambda \cos(x)\right)\Psi(x) = E\,\Psi(x) \,.
\eq
We will refer to the above equation as the quantum Mathieu equation, as in the classical limit $\hbar\ll 1$ we have that
\beq\eqlabel{DopExpansion}
\Dcal\Psi(x)=\Psi(x-\ii\hbar) + \Psi(x+\ii\hbar) = 2 \Psi(x)-\hbar^2\Psi''(x)+\frac{\hbar^4}{12}\Psi''''(x)+\Ocal(\hbar^6)\,,
\eq
and so \req{qMathieuEq} turns at leading order in $\hbar$ into a modified Mathieu equation
\beq\eqlabel{mMathieuEq}
-\Psi''(x)+\frac{2\lambda}{\hbar^2}\cos(x)\Psi(x)+\Ocal(\hbar^2)=\frac{(E-2)}{\hbar^2}\Psi(x)\,.
\eq
The canonical form of the Mathieu equation,
\beq\eqlabel{MathieuEq}
\left(\partial_x^2+\alpha-2q \cos(2x)\right)\Psi(x)=0\,,
\eq
can be obtained via redefining
\beq\eqlabel{MreDefs}
x\rightarrow 2 x\,,\,\,\,\, \lambda\rightarrow \frac{\hbar^2}{4}q \,,\,\,\,\, E\rightarrow 2+\frac{\hbar^2}{4}\alpha\,.
\eq
Note that we have $\lambda$ real and $E>0$ for the region in complex structure moduli space with $z_1 < 0$ real. 

\paragraph{Mathieu energy spectrum}
The perturbative energy spectrum of the Mathieu equation and thus of \req{mMathieuEq} can be obtained for instance by making use of the recursive formula of \cite{FP01}. For $q=\frac{4\lambda}{\hbar^2}\gg 1$, the first few terms of the asymptotic expansion reads
\beq\eqlabel{EfromMathieu}
E = 2(1-\lambda) + (2N+1) \sqrt{\lambda}\hbar-\frac{1+2N+2N^2}{16}\hbar^2-\frac{1+3N+3N^2+2N^3}{256\sqrt{\lambda}}\hbar^3+\Ocal(\hbar^4)\,.
\eq
The expansion can be easily obtained to any desired order (for a brief summary, see appendix B of \cite{K14}). 

The for us here relevant fact is that the asymptotic expansion \req{EfromMathieu} receives non-perturbative corrections, inducing a split of the energy bands $E\rightarrow E^\pm$. This can be verified via computing the true energy spectrum of the classical Mathieu equation numerically, as is briefly sketched in appendix \ref{MathieuNumerics}.

The Mathieu equation completely describes the Nekrasov-Shatashvili limit of four dimensional $\Ncal=2$ $SU(2)$ gauge theory in the $\Omega$-background (see \cite{HM10,K14,BD15} and references therein). The limit $\hbar\rightarrow 0$ keeping 
$$
\alpha=\frac{4(E-2)}{\hbar^2}\,,\,\,\,\, q=\frac{4 \lambda}{\hbar^2}\,,
$$
fixed therefore corresponds to an effective four dimensional field theory limit of \req{qMathieuEq} (in the sense of geometric engineering). Phrased differently, for $\hbar$ sufficiently small, the difference between the energy spectrum of the quantum Mathieu equation \req{qMathieuEq} and the classical Mathieu equation \req{mMathieuEq} becomes negligible. 

In particular, we learned in \cite{K14} (see also \cite{BD15}) that this gauge theory has an intrinsicate non-perturbative structure inherited from the Mathieu equation. For instance, in the regime $q\gg 1$ in moduli space instanton tunneling generate non-perturbative corrections to the quantum periods, measured by an instanton counting parameter $\xi$, which depends on the dynamical scale of the gauge theory (the dynamical scale $\Lambda$ relates to our $\lambda$ parameter as $\lambda\sim\Lambda^2$). The precise instanton counting parameter can be estimated as follows. From exact quantization we know that the instanton action is given by $-\frac{A_p}{\hbar}$ with $A_p$ referring to the leading (perturbative) term of the B-period. The Matone relation \cite{M95} relates $A_p$ to the non-flat coordinate $E$, \ie,
$$
\frac{1}{\hbar}\frac{\partial E}{\partial N} \sim  -c\, \sqrt{\lambda}\frac{\partial A_p}{\partial \sqrt{\lambda}} \,,
$$
with $c$ some constant. More precisely, in the normalization of \cite{K14} we have $c=1/4$. (The origin of the $\hbar$ on the left hand side lies in the quantization condition $a_D=\hbar N$ for the flat coordinate $a_D$.) Hence,
\beq\eqlabel{MathieuXi}
\xi = e^{-\frac{\sqrt{\lambda}}{c\hbar}} = e^{-\frac{\sqrt{q}}{2c}}\,.
\eq
More details can be filled in from \cite{K14}. As it should be, the instanton parameter becomes weaker for large $\lambda$ and vice versa. Therefore we can use the extra parameter $\lambda$ to keep $\xi$ relatively large, while keeping $\hbar$ small.

\paragraph{Quantum Mathieu energy spectrum}
In particular, via tuning $\lambda$ in the quantum Mathieu equation \req{qMathieuEq} we can achieve that
$$
\xi \gg \hbar^2\,,
$$
so that even for very small $\hbar$, where we can compare to the classical Mathieu equation, the non-perturbative corrections are stronger than the ``stringy'' perturbative corrections. With \req{MathieuXi} we infer that this is the case for
\beq\eqlabel{lambdaCond}
\sqrt{\lambda} \ll - \frac{1}{2}\,  \hbar \log\hbar \,.
\eq
One might wonder why we make the effort to obtain the classical Mathieu energy spectrum from \req{qMathieuEq}. The reason is that we can fix in this way a good wave-function basis to expand the solutions of \req{qMathieuEq} into, as the classical results should be reproduced. Naively, as the leading non-constant term in $\hbar$ of \req{EfromMathieu} is given by the quantum harmonic oscillator energy one might think that the harmonic oscillator wave-functions $\Psi$ are a good basis to expand into. For instance such an oscillator basis has been used in \cite{HW14} to calculate the energy spectrum of \req{qMathieuEq} at $\lambda=\ii$ numerically (corresponding to $\lambda=1$ under $x\rightarrow \ii x$). However, this is only part of the story. We learned in section \ref{DefConiSec} that in the complex setting we have in fact two different consistent solutions $\Psi^\pm$, and which one we have to use depends on where we sit in moduli space. However, the precise identification depends on parameterization used. Here, due to the redefinition $x\rightarrow \ii x$, we actually arranged that $\Psi^+$ is consistent for $z_1<0$ and $\Psi^-$ for $z_1>0$. The two solutions $\Psi^\pm$ are fundamentally different, as is as well clear from section \ref{DefConiSec}. One constitutes an expansion into bound states, while the other into resonances. The latter being more interesting from a non-perturbative point of view. 

\begin{figure}[t]
\begin{center}
\psfrag{x}[cc][][1]{$N$}
\psfrag{y}[cc][][1]{$E$}
\includegraphics[scale=0.80]{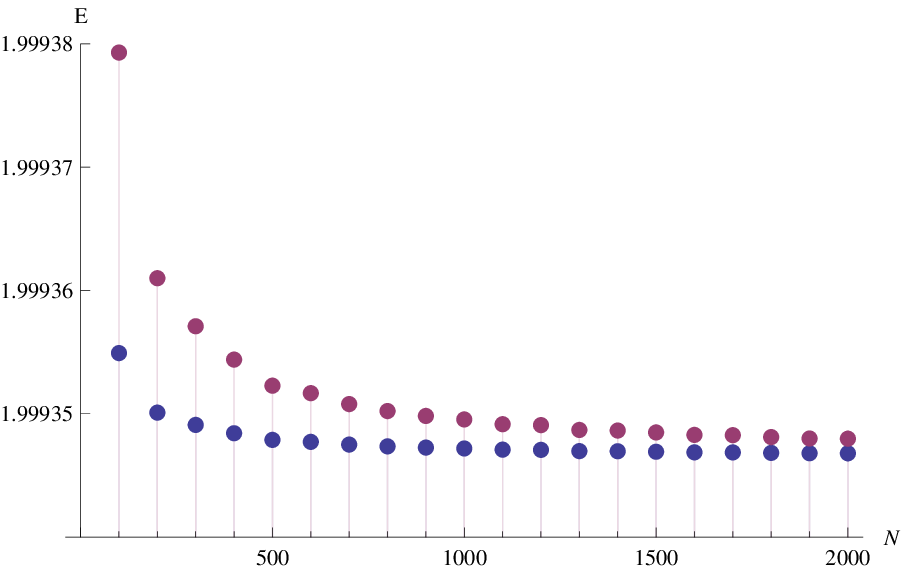}
\includegraphics[scale=0.80]{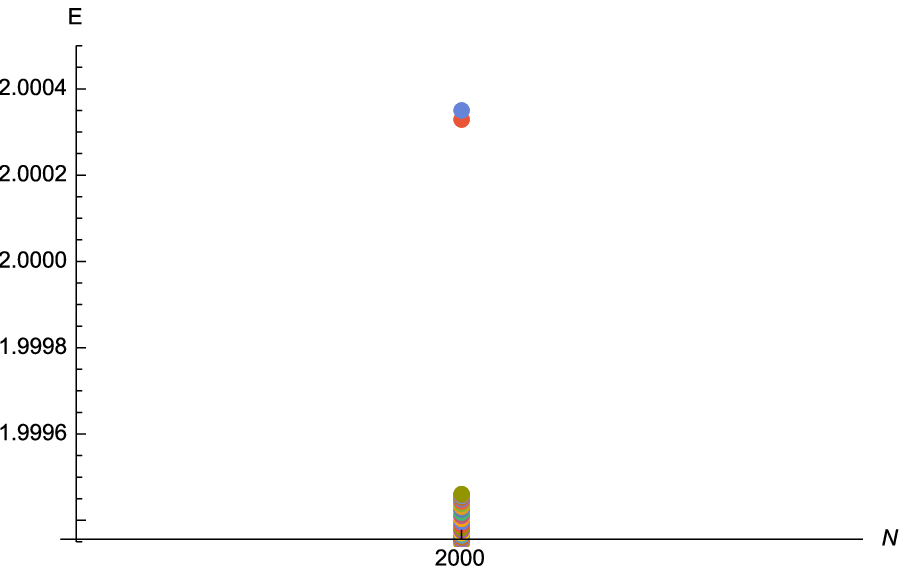}
\caption{Left: The ground (lower blue dots) and first excited state energy of \req{qMathieuEq} for $\hbar=0.05$ and $\lambda=0.001$ as a function of the matrix size $N\times N$ used for numerical evaluation (with step size 100). Right: The first 30 energy levels of \req{qMathieuEq} at $N=2000$}
\label{qMnumConvergenceE0}
\end{center}
\end{figure}

In our parameterization we can easily calculate the energy spectrum of \req{qMathieuEq} numerically. For the readers convenience some more details of the numerical scheme used are recalled in appendix \ref{OsciNumerics}. Note that the rate of convergence is relatively low, for example, we plotted in figure \ref{qMnumConvergenceE0} (left) the ground and first excited state value versus matrix size used in the numerical computation for the point $(\hbar=0.05, \lambda=0.001)$ in parameter space. 

Let us take a more detailed look at this point in parameter space. At matrix size $2000\times 2000$ we obtain $E_0 = 1.999346$, and $E_1=1.999347$ such that $\Delta E_{10}\sim 10^{-6}$. (We only display the first six subdecimal digits.) The string corrections are suppressed by $\hbar^4\sim 6\times 10^{-6}$ (\req{mMathieuEq} has to be multiplied by $\hbar^2$, therefore $\hbar^4$). Hence the observed first two levels actually both correspond to the ground state of the classical Mathieu equation, which we calculate numerically to be $E^c_0=1.999346$. The classical Mathieu equation gives for the first excited state $E^c_1=1.999461$ such that $\Delta E^c_{10}\sim 10^{-4}$ and $\Sigma E^c_{10}=1.999403$, where we defined $\Sigma$ as the mean operator acting on the two energy levels. We have $\hbar^2 e^{-\frac{2\sqrt{\lambda}}{\hbar}}\sim 10^{-4}$ and thus the gap $\Delta E^c_{10}$ is of the expected non-perturbative order. We easily verify that the perturbative energy \req{EfromMathieu} (we are at $q=1.6$) yields for the ground state $E_0^p = 1.999400$, close to $\Sigma E^c_{10}$. 

\begin{figure}[t]
\begin{center}
\psfrag{x}[cc][][1]{$N$}
\psfrag{y}[cc][][1]{$E$}
\includegraphics[scale=0.85]{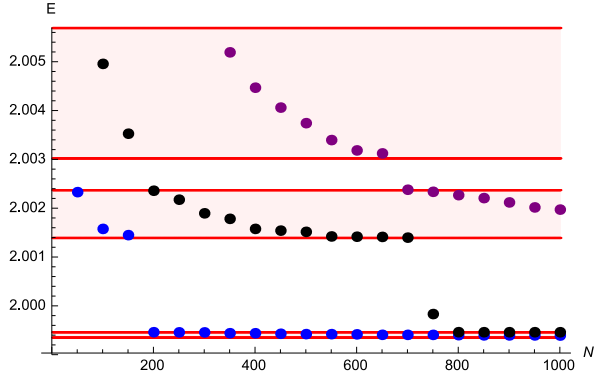}
\caption{\label{qMnumBands}The 8$th$ (lower blue), 16$th$ (middle black) and 30$th$ energy level (top purple dots) of \req{qMathieuEq} at $(\hbar=0.05,\lambda=0.001)$ as a function of $N$ (with step size 50). The first six energy levels of the classical Mathieu equation are indicated by the red lines.}
\end{center}
\end{figure}

It remains to identify the gap $\Delta E^c_{10}$ of non-perturbative origin in the oscillator based numerical energy spectrum $E$. As we saw before for the first two levels, in the gauge theory limit the energy bands collapse (eigenvalues approximately degenerate) and therefore the gaps are sparsely distributed. In particular, the number of eigenvalues degenerating to a particular gauge theory band scales with the matrix dimension $N$. The first 30 energy levels for $N=2000$ are plotted in figure \ref{qMnumConvergenceE0} (right). We observe that there is a diffuse band of 26 eigenvalues around the expected $E^c_1$ and $E^c_2$, followed by a gap with the 27$th$ eigenvalue taking the value $E_{27}=2.001387$. This level corresponds to the third energy level of the classical Mathieu equation, which reads $E_3^c=2.001385$. Hence, it is not easy to resolve the non-perturbative band splitting, \ie, distinguish between $E^c_1$ and $E^c_2$ in the oscillator based numerics. Even so one can tune via the $\lambda$ parameter to a regime in which one can disentangle the non-perturbative contribution from the stringy perturbative corrections (\cf, \req{lambdaCond}), it is difficult to directly resolve the bands, as with increasing $N$ eigenvalues sitting on higher bands fall down to a lower band, thereby diffusing the band structure, as shown in figure \ref{qMnumBands}.
The numerical instability is rooted in the fact that using the $\Psi^+$ wave-function basis essentially corresponds to perturbation around an inharmonic oscillator. Nevertheless, the sharp transitions (jumps) with increasing $N$ can be used to indirectly infer the band structure, as also indicated in figure \ref{qMnumBands}, matching the expectation from the classical Mathieu equation. We conclude that solving \req{qMathieuEq} numerically via expansion in a suitable harmonic oscillator basis indeed reproduces the exact $SU(2)$ gauge theory (classical Mathieu equation) energy spectrum at strong coupling in a suitable decoupling limit.

\begin{figure}[t]
\begin{center}
\psfrag{x}[cc][][1]{$N$}
\psfrag{y}[cc][][1]{$E$}
\includegraphics[scale=0.85]{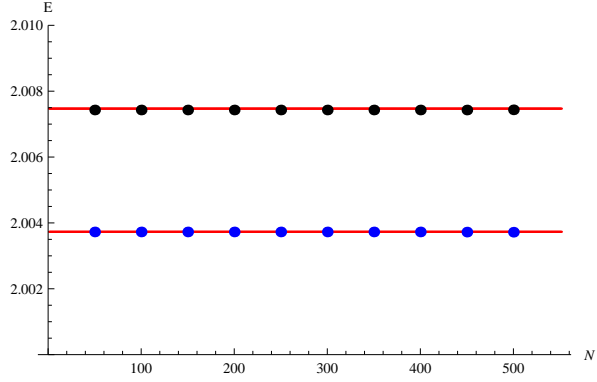}
\caption{The 1$th$ (lower blue) and 2$nd$ (black) energy level of \req{qMathieuEq} at $(\hbar=0.05,\lambda=0.001)$ solved via the $\Psi^-$ basis as a function of $N$ (with step size 50). The first two energy levels of the perturbative solution of the classical Mathieu equation (under $\hbar\rightarrow -\ii\hbar,\lambda\rightarrow-\lambda$) are indicated by the red lines.}
\label{qMnumBands2}
\end{center}
\end{figure}

In comparison, let us observe what we would have obtained if we use instead the other basis $\Psi^-$. As described in the appendix, in the numerical evaluation one can rotate between $\Psi^\pm$ simply via reparameterizing the curve $\Sigma$ by $x\rightarrow \ii x$. Hence, this corresponds to the numerics invoked in \cite{HW14}. We plot the first two obtained energy levels at the same point $\hbar=0.05,\lambda=0.001$ in figure \ref{qMnumBands2}. 
For instance, we obtain for the ground state $E_0=2.003717$, which matches the perturbative energy $E^p_0=2.003725$ obtainable from \req{EfromMathieu} under $\hbar\rightarrow -\ii\hbar$ and $\lambda\rightarrow -\lambda$ up to stringy corrections of order $\hbar^4$. In particular, we do not observe band splitting.

\paragraph{An Explanation}

Let us give an explanation why the numerics performed works and what it actually calculates. Consider the curve \req{clCurve} and rescale coordinates by a small factor $l$. This yields,
\beq\eqlabel{F0energyExpansion}
E=2(1+\lambda) + (p^2-\lambda x^2)l^2+\Ocal(l^3)\,.
\eq
Hence, close to the origin the geometry is described at leading order by a deformed conifold. In particular, the discussion of section \ref{DefConiSec} applies. 

Note that in the geometry \req{F0origCurve} we can introduce local coordinates $\Delta_i$ near a conifold point via (see for example \cite{HKR08})
\beq\eqlabel{F0coniCord}
z_1 = \frac{1}{8}-\frac{1}{8(2+\Delta_1(\Delta_2-1)-\Delta_2)}\,,\,\,\,\,\, z_2=\frac{\Delta_2-1}{8(2+\Delta_1(\Delta_2-1)-\Delta_2)} \,. 
\eq
For simplicity, let us consider the case with $\lambda=1$ of \cite{HW14}. This translates to $\Delta_1=0$, and we denote the remaining modulus simply as $\Delta$. The map to the $E$ coordinate of the curve \req{clCurve} reads then
\beq\eqlabel{EF0viaDelta}
E=2\sqrt{\frac{2(\Delta-2)}{\Delta-1}}=4+\Delta +\Ocal(\Delta^2)\,.
\eq
Comparing with \req{F0energyExpansion}, we deduce that the numerics actually calculates the mirror map at the conifold point in moduli space. Let us verify this explicitly. 

Inserting the coordinates \req{F0coniCord} into \req{F0origCurve} (and rotating for convenience $x\rightarrow\ii x$) we obtain the difference equation
$$
(2-\Delta+\frac{1}{8}( \Delta-1)e^{-x}+(\Delta-2)e^x)\Psi(x)+ (\Delta-2) \Psi(x+\hbar)+\frac{1}{8}(\Delta-1)\Psi(x-\hbar)=0\,.
$$
We can solve the difference equation perturbatively via a WKB Ansatz. That is, we set
$$
\Psi(x) = e^{\frac{S(x;\hbar)}{\hbar}}\,,
$$
with 
$$
S(x;\hbar) = \sum_{n=0}^\infty S^{(n)}(x)\, \hbar^n \,,
$$
and solve for $\partial_x S^{(n)}$ order by order in $\hbar$. Expanding for small $\Delta$ and taking residue at $-\log 4$ we obtain a (perturbative) quantum flat coordinate
\beq\eqlabel{tcDef}
t_c = \sum_{n=0}^\infty t_c^{(n)}\,\hbar^n\,,
\eq
with the first few 
\beq\eqlabel{tcnDef}
t_c^{(n)}= {\rm Res} \, \partial_x S^{(n)}\,,
\eq
reading 
\beq
\begin{split}
t_c^{(0)}&=-\ii \left(\frac{1}{2} \Delta+\frac{13}{32}\Delta^2+\frac{521}{1536}\Delta^3+\frac{4749}{16384}\Delta^4+\Ocal(\Delta^5) \right)\,,\\
t_c^{(1)}&=-\frac{1}{2}\,,\\
t_c^{(2)}&=\ii\left(-\frac{1}{32}+\frac{1}{512}\Delta+\frac{9}{8192}\Delta^2+\frac{165}{262144}\Delta^3+\frac{3091}{8388608}\Delta^4+\Ocal(\Delta^5) \right)\,,\\
t_c^{(3)}&=0 \,,\\
t_c^{(4)}&= \ii\left(\frac{13}{49152}-\frac{275}{1572864}\delta-\frac{597}{8388608}\Delta^2-\frac{3517}{134217728}\Delta^3 +\Ocal(\Delta^4) \right)\,.
\end{split}
\eq
Note the non-vanishing $t_c^{(1)}$. Inverting the WKB flat-coordinate $t_c$ yields
\beq
\begin{split}
\Delta^{(0)}_p &= \ii2 t_c+\frac{13}{4}t_c^2-\ii\frac{493}{96}t_c^3-\frac{12427}{1536}t_c^4+\Ocal(t_c^5)\,,\\
\Delta^{(1)}_p &= \frac{1}{2} \partial_{t_c}\Delta^{(0)}_p\,,\\ 
\Delta^{(2)}_p &= \frac{3}{4}-\ii\frac{233}{64} t_c-\frac{93}{8} t_c^2-\ii \frac{756253}{24576}t_c^3-\frac{211675}{98304}t_c^4+\Ocal(t_c^5)\,,\\
\Delta^{(3)}_p &= -\ii\frac{103}{192}-\frac{5429}{1536}t_c-\ii\frac{351443}{24576}t_c^2-\frac{211675}{49152}t_c^3+\Ocal(t_c^4) \,,\\
\Delta^{(4)}_p &= -\frac{195}{512}+\ii\frac{156421}{49152}t_c-\frac{5005231}{1572864}t_c^2+\ii\frac{5756235 }{8388608} t_c^3+\Ocal(t_c^4) \,.
\end{split}
\eq
Invoking the usual Bohr-Sommerfeld quantization for the A-period, \ie, $t_c= \hbar N$, we obtain (we also rotated $\hbar\rightarrow -\ii\hbar$)
\beq\eqlabel{F0DeltaN}
\begin{split}
\Delta_p(N)=&\, (2N+1)\hbar-\frac{3+13N+13N^2}{4}\hbar^2+\frac{103+699N+1479N^2+986N^3}{192}\hbar^3\\
&-\frac{585+5429N+17856N^2+24854N^3+12427N^4}{1536}\hbar^4+\Ocal(\hbar^5)\,.
\end{split}
\eq
We have now everything at hand to compare the perturbative WKB mirror map against the numerical computation in the harmonic oscillator phase. We list the obtained $E(\Delta_p(0))$ and numerical $E_0$ for various $\hbar$ in table \ref{F0Etab}. We infer that the perturbative WKB energies appear to converge with increasing WKB order to the numerical $E_0$. More precisely, up to Borel resummation, as has been observed and discussed before in \cite{H15} via a high order WKB calculation of \req{F0energyExpansion}.\footnote{We like to thank Y. Hatsuda for related discussions.}
\begin{table}[t]
\begin{center}
\begin{tabular}{c|c|c|c}
$\hbar$ &$E(\Delta_p(0)+\Ocal(\hbar^4))$& $E(\Delta_p(0)+\Ocal(\hbar^6))$&    $E_0$\\
\hline
$0.10$ & \underline{4.101255}2083 & \underline{4.101255335}8 & 4.1012553359 \\
$0.25$ & \underline{4.25789}38802 & \underline{4.2578987}002 & 4.2578987246 \\
$0.50$ & \underline{4.5319}010416 & \underline{4.53197}39024 & 4.5319753251 \\ 
$0.75$ & \underline{4.822}5097656 & \underline{4.8228}570580 & 4.8228719839 \\ 
$1.00$ & \underline{5.13}02083333 & \underline{5.131}2377929 & 5.1313156016 \\
$1.25$ & \underline{5.45}54850260 & \underline{5.45}78319589 & 5.4581090443 \\
$1.50$ & \underline{5.}7988281250 & \underline{5.80}33496856 & 5.8041260743 \\
\end{tabular}
\end{center}
\caption{\label{F0Etab}The perturbative energies $E(\Delta_p(0))$ for various $\hbar$ obtained via \req{EF0viaDelta} from the conifold coordinate \req{F0DeltaN} versus the numerical result in the harmonic oscillator basis for matrix size $200\times200$. Matching digits are underlined. We show only the leading 10 subdecimal digits.}
\end{table}
As a side remark, note that for increasing $N$ we are moving away from the conifold point and the expansion \req{F0DeltaN} starts to break down. Similarly, the numerical approximation becomes less and less accurate with increasing energy level.

Let us consider now the other phase, that is where we expand into an inharmonic oscillator basis. We set $\lambda=-1$ to reach this phase. From \req{F0energyExpansion} we learn that in this case 
$$
E=\Delta\,. 
$$
However, $\Delta$ is not the same conifold coordinate as above, because \req{F0coniCord} is not a good coordinate around $z_1=-z_2$. Instead, we directly solve \req{qMathieuEq} via a WKB Ansatz, since this will directly yield the appropriate $\Delta$ via taking residue at zero.

We obtain for the quantum A-period expansion
\beq
\begin{split}
t_c^{(0)}&=-\frac{1}{2}\Delta-\frac{1}{384}\Delta^3-\frac{9}{163840}\Delta^5 +\Ocal(\Delta^7)\,, \\
t_c^{(1)}&=-\frac{1}{2}\,,\\
t_c^{(2)}&=-\frac{1}{128}\Delta-\frac{5}{16384}\Delta^3-\frac{35}{2097152}\Delta^5 +\Ocal(\Delta^7) \,, \\
t_c^{(3)}&=0\,,\\
t_c^{(4)}&=-\frac{19}{98304}\Delta -\frac{223}{6291456}\Delta^3-\frac{14245}{3221225472}\Delta^5+\Ocal(\Delta^7) \,.
\end{split}
\eq
Inversion yields,
\beq
\begin{split}
\Delta^{(0)}_p &= -2t_c+\frac{1}{24} t_c^3+\frac{7}{7680} t_c^5+\Ocal(t_c^7)\,,\\
\Delta^{(1)}_p &= \frac{1}{2}\partial_{t_c} \Delta^{(0)}_p\,,\\ 
\Delta^{(2)}_p &= \frac{1}{16}t_c+\frac{7}{1536}t_c^3+\frac{203}{491520}t_c^5+\Ocal(t_c^7)\,,\\
\Delta^{(3)}_p &= \frac{1}{48}+\frac{7}{1536}t^2+\frac{203}{196608}t_c^4  +\Ocal(t_c^6)\,,\\
\Delta^{(4)}_p &= \frac{7}{3072}t_c+\frac{1571}{1179648}t_c^3+\frac{107287}{754974720}t_c^5+\Ocal(t_c^7)\,.\\
\end{split}
\eq
Quantization of the period via $t_c= \hbar N $ leads to 
\beq\eqlabel{F0DeltaN2}
\begin{split}
-\Delta_p(N) = &\, (2N+1)\hbar -\frac{1+3N+3N^2+2N^3}{48}\hbar^3 \\ 
& +\frac{7(1+5N+10N^2+10N^3+5N^4+2N^5)}{15360}\hbar^5 +\Ocal(\hbar^7) \,.
\end{split}
\eq

The first two energy levels obtained numerically in the $\Psi^-$ phase are listed for various $\hbar$ in table \ref{F0Etab2}. 
\begin{table}[t]
\begin{center}
\begin{tabular}{c|c|c|l}
$\hbar$ &$E(\Delta_p(0)+\Ocal(\hbar^6))$&   $E^\pm_0$ & $\sim\Delta E_{10}$\\
\hline
$0.25$ & \underline{0.24967403}41  & $\begin{matrix}\underline{0.24967403015556}\\ \underline{0.24967403015556}\end{matrix}$ & $< 10^{-15}$ \\
\hline
$0.50$ & \underline{0.497381}5917 &  $\begin{matrix}\underline{0.497381}01940237 \\ \underline{0.497381}10400013 \end{matrix}$& $9\times 10^{-8}$ \\
\hline 
$0.75$ & \underline{0.741}1027908 & $\begin{matrix}\underline{0.741}06083750997 \\ \underline{0.741}12122095652 \end{matrix}$& $6\times 10^{-5}$ \\
\hline 
$1.00$ & \underline{0.97}87109375 & $\begin{matrix}\underline{0.97}735708912389 \\ \underline{0.97}814526674647 \end{matrix}$& $8\times 10^{-4}$  \\
\hline
$1.25$ & \underline{1.}2079191207 & $\begin{matrix} \underline{1.}19868743160450 \\ \underline{1.}20405179498400 \end{matrix}$& $5\times 10^{-3}$ \\
\hline
$1.50$ & \underline{1.}4262268066 & $\begin{matrix} \underline{1.}39390178777156\\ \underline{1.}41203707527623\end{matrix}$& $2\times 10^{-2}$  \\
\end{tabular}
\end{center}
\caption{\label{F0Etab2}The perturbative WKB energy (sign flipped), the first two numerical energy levels via the inharmonic oscillator basis and the resulting band widths for matrix size $100\times100$. The numerical energy correspond to a non-perturbative completion of the quantum mirror map \req{F0DeltaN2} of local $\P^1\times \P^1$ at $z_2\rightarrow \infty$ and $\lambda=-1$. Matching digits are underlined.}
\end{table}
We observe non-perturbative band splitting, with the width of bands scaling exponentially, as expected from our discussions in section \ref{DefConiSec}. However, one should keep in mind that the obtained values for $E^\pm_0$ and $\Delta E_{10}$ are only qualitative in nature due to the intrinsic instability of the numerics (\cf, figure \ref{qMnumBands} and previous discussions).\footnote{Precise results can be obtained via a different numerical scheme, confirming the qualitative results given in table \ref{F0Etab2}. Details will appear elsewhere.}

Note that the perturbative WKB expansion is not able to resolve the bands, but rather 
$$
E(\Delta_p(0)) \sim \Sigma E_{10}\,,
$$
with increasing accuracy for more WKB orders taken into account. Hence,  
$$
E_0 = E(\Delta_p(0)) \pm \Ocal(\xi)\,,
$$
where $\xi$ denotes an instanton counting parameter. We conclude that in this phase the WKB expansion is non-perturbatively corrected, in contrast to the phase discussed before. This confirms statements about the non-perturbative completion of the NS limit of topological strings made in the first part of this series \cite{K13}.

\section{Local $\P^2$}
\label{P2sec}
The classical curve in a parameterization convenient to extract the large volume periods at $z\ll 1$ reads
\beq\eqlabel{P2curve}
\Sigma: -1+e^x+e^p+ze^{-x-p}=0\,.
\eq
We redefine similar as in \cite{HW14}
$$
z\rightarrow \frac{1}{E^3}\,,\,\,\,\,\, x\rightarrow x-\log E\,,\,\,\,\,\,p\rightarrow p-\frac{x}{2}-\log E\,,
$$
yielding the curve
\beq\eqlabel{P2curve2}
e^x+e^{-\frac{x}{2}+p}+e^{-\frac{x}{2}-p}=E\,.
\eq
Note that under this redefinition the large volume regime is now located at $E\gg 1$.

The geometry develops a conifold singularity at $z=\frac{1}{27}$ and we introduce a local coordinate $\Delta$ as
$$
z=\frac{1-\Delta}{27}\,.
$$
Hence, we have a simple map between the $E$ and $\Delta$ coordinates
\beq\eqlabel{EviaDelta}
E = \frac{3}{(1-\Delta)^{1/3}}=3+\Delta+\Ocal(\Delta^2)\,.
\eq
Note that the above map can also be inverted. The inverse series for $\widetilde E:= (E-3)/3$ reads
$$
\Delta(\widetilde E) = \frac{1}{2}\sum_{n=2}^\infty (-1)^n\, n(n+1) \, \widetilde E^{n-1}\,.
$$

Rescaling in \req{P2curve2} the coordinates by $l$ and expanding for small $l$ yields
\beq\eqlabel{P2Eexpand}
E=3+\left(p^2+\frac{3x^2}{4}\right)l^2+\Ocal(l^3)\,.
\eq
Hence, close to the origin the geometry corresponds in the parameterization \req{P2curve2} at leading order to a deformed conifold, similar as in the previous section. In particular, we infer from \req{EviaDelta} that the complex structure modulus thereof reads $\Delta$. 

The curve \req{P2curve} can be quantized as usual. However, one should keep in mind that according to the above discussion the ground state sits close to the conifold point in moduli space (and can be mapped thereto via \req{EviaDelta}). This explains why one can numerically approximate the quantum energies via expansion into an oscillator basis. In particular, the discussion of section \ref{DefConiSec} applies.

In order to infer the perturbative quantum periods at the conifold point in moduli space, it is more convenient to directly parameterize the curve \req{P2curve} in terms of $\Delta$, yielding 
$$
1-27 (1-e^{x})e^{x+p}+27 e^{x+2p}=\Delta\,.
$$
Expanding $p(\Delta)$ for small $\Delta$, we observe that $p$ has a pole at $x_*=-\log 3$. Taking residue, yields
$$
{\rm Res}\,  p(\Delta) = -\frac{\ii}{\sqrt{3}}\left( \Delta+\frac{11}{18} \Delta^2+\frac{109}{243}\Delta^3+\frac{9389}{26244}\Delta^4+\frac{88351}{295245}\Delta^5+\Ocal(\Delta^6)\right)\,.
$$
Denoting the flat coordinate at the conifold point as $t^{(0)}_c$, we infer that (see for instance \cite{KW09})
$$
t^{(0)}_c = 3\ii \, {\rm Res}\,  p(\Delta)\,.
$$
The perturbative quantum geometry, as reviewed in section \ref{QGsec}, arises via canonical quantization. For $[x,p]=-\hbar$, and making use of the Baker-Campell-Hausdorff formula (\cf, \req{BCHformula}), we obtain the quantum curve
\beq\eqlabel{P2coniDiffEq}
(1-\Delta)\,\Psi(x) - 27 e^{\frac{\hbar}{2}} e^x(1-e^{\frac{\hbar}{2}}e^x)\, \Psi(x+\hbar)+27e^{\hbar} e^{x}\, \Psi(x+2\hbar) =0\,.
\eq
Performing a WKB Ansatz for $\Psi$ and expanding for $\Delta$ small, we obtain similar as in the previous section for the periods \req{tcnDef}
\beq\eqlabel{P2tExpand}
\begin{split}
t_c^{(1)} & =-\frac{3\ii}{2} \,, \\
t_c^{(2)} & = -\sqrt{3} \left(\frac{1}{36} +\frac{1}{324}\Delta+\frac{5}{4375}\Delta^2+\frac{35}{59049}\Delta^3+\frac{385}{1062882}\Delta^4+\Ocal(\Delta^5)\right)\,,\\
t_c^{(3)} & = 0 \,, \\
t_c^{(4)} & = -\sqrt{3}\left(\frac{19}{139968}-\frac{91}{1259712}\Delta-\frac{89}{2834352}\Delta^2-\frac{3521}{229582512}\Delta^3+\Ocal(\Delta^4)\right)  \,.\\
\end{split}
\eq
($t_c^{(n>1)}$ vanishes for $n$ odd.) Up to overall normalization, and the constant at order $\hbar^1$, the expansions given above are in accord with \cite{HKRS14}, where the higher order $t_c^{(n)}$ have been obtained by acting with certain differential operators on $t_c^{(0)}$, similar as has been done before in the case of the periods of the Seiberg-Witten curve \cite{MM09} and Dijkgraaf-Vafa geometries \cite{ACDKV11}. The non-vanishing term of order $\hbar^1$ is however quite important.

Inverting the WKB flat-coordinate $t_c$ as given through \req{tcDef} and \req{tcnDef}, yields the perturbative coordinate $\Delta_p(t_c)$. The first few orders in $\hbar$ read
\beq
\begin{split}
\Delta_p^{(0)} &= -\frac{1}{\sqrt{3}}t-\frac{11}{64} t_c^2-\frac{145}{1458\sqrt{3}}t_c^3-\frac{6733}{472392}t_c^4+\Ocal(t^5)  \,,\\
\Delta_p^{(1)} &= \frac{3\ii}{2} \partial_{t_c}\Delta_p^{(0)} \,,\\
\Delta_p^{(2)} &= \frac{31}{72}+\frac{415}{648\sqrt{3}} t_c+\frac{19487}{104976}t_c^2+\frac{116831}{944784\sqrt{3}}t_c^3+\Ocal(t^4)\,,\\
\Delta_p^{(3)} &= \frac{125\ii}{432\sqrt{3}}+\frac{223\ii}{1296}t_c +\frac{110239\ii}{629856\sqrt{3}} t_c^2-\frac{367\ii}{118098}t_c^3+\Ocal(t^4) \,,\\
\Delta_p^{(4)} &= -\frac{16073}{279936}  -\frac{302785t}{2519424\sqrt{3}}t_c+\frac{115931}{17006112}t_c^2-\frac{9126013}{2754990144\sqrt{3}}t_c^3+\Ocal(t^4) \,.\\
\end{split}
\eq
Note that due to the non-vanishing order $\hbar^1$ in \req{P2tExpand} the quantum coordinate $\Delta_p$ will be a series in even and odd powers of $\hbar$. Under quantizing $t_c= 3\ii \hbar\, N$ and rotating $\hbar\rightarrow \ii\hbar$ we obtain
\beq\eqlabel{DeltaN}
\begin{split} 
\Delta_p(N) =& \frac{\sqrt{3}(2N+1)}{2}\hbar-\frac{31+132N+132N^2}{72}\hbar^2+\frac{5(25+166N+348N^2+232N^3)}{432\sqrt{3}}\hbar^3 \\
&-\frac{16073+144504N+467688N^2+646368N^3+323184N^4}{279936}\hbar^4+\Ocal(\hbar^{5})\,.
\end{split}
\eq

We can now compare $\Delta_p(0)$ against the ground state $E_0$ obtained via numerically approximating the energies of the quantum curve resulting from \req{P2curve2} under quantizing $[x,p]=\ii\hbar$ and using the $\Psi^+$ basis with $\kappa = \frac{\sqrt{3}}{2\hbar}$ (\cf, appendix \ref{OsciNumerics}). The results are listed in table \ref{P2Etab}. 
\begin{table}[t]
\begin{center}
\begin{tabular}{c|c|c|c}
$\hbar$ &$E(\Delta_p(0)+\Ocal(\hbar^4))$& $E(\Delta_p(0)+\Ocal(\hbar^6))$ &   $E_0$\\
\hline
$0.10$ & \underline{3.0873036}671 & \underline{3.0873036489}  & 3.0873036489\\
$0.25$ & \underline{3.22095}10395 & \underline{3.22095037}52  & 3.2209503734\\
$0.50$ & \underline{3.451}2090998 & \underline{3.451199}6751  & 3.4511995539 \\ 
$0.75$ & \underline{3.691}4006459 & \underline{3.69135}90292  & 3.6913576372 \\ 
$1.00$ & \underline{3.942}1521430 & \underline{3.94203}98804  & 3.9420320545\\
$1.25$ & \underline{4.20}40900561 & \underline{4.2038}630151  & 4.2038333141 \\
$1.50$ & \underline{4.477}8408505 & \underline{4.477}4675954  & 4.4773797291 \\
\end{tabular}
\end{center}
\caption{\label{P2Etab} The perturbative energies $E(\Delta_p(0))$ for various $\hbar$ obtained via \req{EviaDelta} from the conifold coordinate \req{DeltaN} versus the numerical result in the oscillator basis for matrix size $300\times300$. Matching digits are underlined. We show only the leading 10 subdecimal digits.}
\end{table}
We infer that the perturbative $E(\Delta_p(0))$ converges with increasing WKB order to the numerical $E_0$ (up to Borel resummation, \cf, \cite{H15}). Hence, as is already clear from \req{EviaDelta} and \req{P2Eexpand}, the numerical energy $E_0$ in the $\Psi^+$ (harmonic oscillator) phase just corresponds to the conifold mirror map, which in this case does not receive non-perturbative corrections.

The from a non-perturbative perspective actually interesting case corresponds to the other possible wave-function solution $\Psi^-$ obtainable via a suitable parameterization of the curve, as is clear from \req{P2Eexpand}. According to our discussions in section \ref{DefConiSec} we expect a non-trivial non-perturbative structure to be present in this phase. The reason is that in this case we have a parabolic barrier which is cutoff (regularized) by the embedding into the Calabi-Yau. However, things are technically more involved as in our previous discussions. Firstly, $c_X \sim \frac{\pi}{3}+1.678699904\ii$ is complex valued \cite{HKR08}, and so will be the instanton counting parameter $\xi$. In particular, this implies that the resulting energy bands have an imaginary part. Secondly, because we do not have another moduli to rotate as for $\P^1\times\P^1$ discussed before, we have to rotate instead $\hbar\rightarrow \ii\hbar$ in order to obtain a consistent $\Psi^-$ solution. However, this solution sits at sign flipped energies. In particular, the energy decreases (and turns negative) with increasing energy level. This obscures the numerics in a similar way as shown in the previous section in figure \ref{qMnumBands}. However, in the current case things are inverted, that is the eigenvalues flow down from the ground state with increasing matrix dimension, and not towards it as we had before. Nevertheless, we can still perform a qualitative check. For instance at $\hbar=-0.5$ we obtain from \req{DeltaN} the first few digits $\Delta_p(0)=2.583498$. The numerics at matrix dimension $200\times 200$ yields for the groundstate (now corresponding to the highest energy level) $E_0= 2.583503+\ii 8.280180\times 10^{-6}$. Hence,
$$
(E_0-\Delta_p(0))|_{\hbar=0.5} \sim 10^{-5} - \ii 10^{-5}\,, 
$$
and indeed we have a small real and imaginary perturbation away from the perturbative WKB solution. Inspection of the eigenvalue distributions for various $\hbar$ suggest that the magnitute of the perturbation scales exponentially, as is implied by section \ref{DefConiSec}. However, as the eigenvalues in the numerics are diffused in both the real and imaginary direction, we refrain here to perform a more quantitative analysis. We leave this topic for another research project, perhaps making use of some better behaved numerical scheme to solve the difference equation at the point in moduli space of interest.

\appendix

\section{Numerics}
\label{NumericSec}
\subsection{Numerical Mathieu spectrum}
\label{MathieuNumerics}
We compute the energy spectrum of \req{mMathieuEq} numerically following \cite{CVY09}. According to Floquet's theorem, there exists always a solution to Mathieu's equation \req{MathieuEq} of the form 
$\Psi(x) = e^{\ii \nu(\alpha,q) x}\ \varpi(x)$\,,
where $\nu$ is called characteristic exponent and $\varpi(x)$ some periodic function of period $\pi$. We can distinguish different solutions by the values $\nu$ takes. For us of relevance is $\nu=N\in\Z$, corresponding to bounded and periodic (in $\pi$ or $2\pi$) solutions. These are the Mathieu functions of the first kind. Clearly, for $\nu$ integer, $\Psi(x)$ is periodic with period $\pi$ or $2\pi$ (depending on if $N$ is even or odd) and therefore we can insert a Fourier Ansatz into \req{MathieuEq} to obtain a recurrence relation for the expansion coefficients $c_k$, given by the matrix equation $(M-\alpha \mathbb I)c=0$ with
\beq
M=
\left(
\begin{matrix}
\ddots& & & & & & & &\\
& 3^2 &  & q &  &  &  &  &\\
&  & 2^2 &  & q &  &  &  &\\
& q &  & 1^2 &  & q &  &  &\\
&  & q &  &  0^2 &  & q &   &\\
&  &  & q &  & 1^2 &  & q &\\ 
&  &  &  & q &  & 2^2 &  &\\
&  &  &  &  & q &  & 3^2 & \\ 
& & & & & & & &\ddots
\end{matrix}
\right)\,.
\eq
This linear system has non-trivial solutions if $\det(M-\alpha \mathbb I)=0$. Hence, the energy spectrum $E$ of \req{mMathieuEq} is approximated by the eigenvalues of $M$, \ie,
$$
E(N) = 2+\frac{\hbar^2}{4} {\rm Ev}(M)|_N\,,
$$
where we made use of the relation \req{MreDefs} and ${\rm Ev}(M)|_N$ refers to the $Nth$ eigenvalue of $M$, ordered in increasing order.

\subsection{Oscillator basis expansion}
\label{OsciNumerics}
Recall from section \ref{DefConiSec} that the wave-functions
$$
\Psi_n(x;\kappa) = \frac{1}{\sqrt{2^n n!}} \left(\frac{\kappa}{\pi}\right)^{1/4} e^{-\frac{\kappa x^2}{2}} H_n(\sqrt{\kappa} x)\,,
$$
form a solution basis for the complex square potential operator. The real section (quantum mechanics in 1d) interpretation of the wave-functions depends on the particular value $\kappa$ takes in the complex plane. In particular, for $\kappa$ real and positive these wave-functions form the well-known basis of the quantum harmonic oscillator to positive energy. In the complex plane we should distinguish between the two solutions $\Psi^\pm$ related to each other via $\kappa\rightarrow -\kappa$, as the integration contour has to be taken either along the real or imaginary axis. Therefore, we have two different matrix elements $M^\pm$,
\beq\eqlabel{MatrixElement}
M^\pm_{n_1,n_2}=\bra{\Psi^\pm_{n_1}} f(e^x,e^p) \ket{\Psi^\pm_{n_2}} = \int_{\Ccal_\pm} dx\,, \cc{\Psi_{n_1}(\kappa;x)} f(e^x,e^p)\Psi_{n_2}(x;\kappa)  \,,
\eq
with $f$ some polynomial in $e^x$ and $e^p$. (Depending on the geometry $f$ there can also be more inequivalent matrix elements.) Let us first consider an expansion into the $\Psi^+$ basis (we fix $\kappa$ such that $|\Re\kappa| >|\Im\kappa|$) and proceed as in \cite{HW14}. In order to evaluate \req{MatrixElement}, we make use of the integral (for $n\leq m$)
$$
\int_{-\infty}^\infty dx\, e^{-x^2} H_{n}(x+y)H_{m}(x+z) = 2^{m}\sqrt{\pi} n!\,z^{m-n} L_{n}^{(m-n)}(-2yz)\,,
$$
with $L^{(\alpha)}_n$ the Laguerre polynomials. We infer
\beq
\begin{split}
\bra{\Psi^+_{n_1}}e^{\frac{ab}{2}} e^{ax +b\partial_x} \ket{\Psi^+_{n_2}} =&\,\sqrt{\frac{2^{n_2}\,n_1!}{2^{n_1}\,n_2!}} \, \left(\frac{a+b\kappa}{2\sqrt{\kappa}}\right)^{n_2-n_1} e^{\frac{a^2-b^2\kappa^2}{4\kappa}}\\
&\times  L^{(n_2-n_1)}_{n_1}\left(\frac{b^2\kappa^2-a^2}{2\kappa}\right) \,,
\end{split}
\eq
and hence $M^+$ can be easily calculated up to some desired matrix size. Note that we included the additional factor $e^{\frac{a b}{2}}$ in the matrix element as such a factor arises under quantization from the Baker-Campell-Hausdorff formula,
\beq\eqlabel{BCHformula}
e^{X}e^Y=e^{X+Y+\frac{1}{2}[X,Y]}\,,
\eq
for central commutator. The energy spectrum follows as usual via calculating the eigenvalues, \ie,
$$
E(N) = {\rm Ev}(M)|_N\,,
$$
where the set of eigenvalues is taken to be ordered in increasing order as in appendix \req{MathieuNumerics}.

It remains to discuss the case with $\Psi^-$ ($|\Re\kappa| < |\Im\kappa|$). We take as integration contour in this case $\Ccal_-=[-\ii\infty,\ii\infty]$. Note that we can rotate $\Ccal_-$ to $\Ccal_+$ via $x\rightarrow -\ii x$ and can absorb the gained $-\ii$ in the wave-functions via sending $\kappa\rightarrow -\kappa$. Hence,
$$
M^-_{n_1,n_2} = \bra{\Psi^+_{n_1}}f(e^{-\ii x},e^p)\ket{\Psi^+_{n_2}}\,,
$$
which can be evaluated as above. Note that the rotation of contour is performed after the operator $p$ acts on the wave-function, therefore the quantization condition does not change.

\end{document}